\documentclass[aps,prd,preprint,showpacs,superscriptaddress]{revtex4}

\usepackage{amstext}

\usepackage{graphicx}
\usepackage{amsmath}
\usepackage{amssymb}
\usepackage{dcolumn}
\usepackage{bm}

\begin{document}

\title{Model description of non-Maxwellian nuclear processes in the solar interior}

\author{Victor~T.~Voronchev}
\affiliation{Skobeltsyn Institute of Nuclear Physics, Lomonosov Moscow State University, Moscow 119991, Russia}
\email[]{voronchev@srd.sinp.msu.ru}

\author{Yasuyuki~Nakao}
\affiliation{Green Asia Education Center, Kyushu University, 6-1 Kasuga-koen, Kasuga, Fukuoka 816-8580, Japan}
\email[]{nakao@nucl.kyushu-u.ac.jp}

\author{Yukinobu~Watanabe}
\affiliation{Department of Advanced Energy Engineering Science, Kyushu University, 6-1 Kasuga-koen, Kasuga, Fukuoka 816-8580, Japan}
\email[]{watanabe@aees.kyushu-u.ac.jp}


\begin{abstract}
A consistent model for the description of non-Maxwellian nuclear processes in the solar core triggered by fast reaction-produced particles is formulated. It essentially extends an approach to study suprathermal solar reactions discussed previously [Phys. Rev. C \textbf{91}, 028801 (2015)] and refines its predictions. The model is applied to examine in detail the slowing-down of 8.7-MeV $\alpha$ particles produced in the $\mathrm{^{7}Li}(p,\alpha)\alpha$ reaction of the $pp$ chain, and to study suprathermal processes in the solar carbon-nitrogen-oxygen (CNO) cycle induced by them. The influence of electron degeneracy and electron screening on suprathermal reactions through in-flight reaction probability and fast particle emission rate is clarified. In particular, these effects account for a 20\% increase of the $\mathrm{^{14}N}(\alpha ,p)\mathrm{^{17}O}$ reaction rate at $R < 0.2 R_\odot$. This new type of correction is important for the suprathermal reaction like $\mathrm{^{14}N}(\alpha ,p)\mathrm{^{17}O}$ as it is recognized to be capable of distorting the CNO cycle in the 95\% region of the solar core. In this region, normal branching $\mathrm{^{14}N} \leftarrow \mathrm{^{17}O} \rightarrow \mathrm{^{18}F}$ of nuclear flow transforms to abnormal sequential flow $\mathrm{^{14}N} \rightarrow \mathrm{^{17}O} \rightarrow \mathrm{^{18}F}$, and the $\mathrm{^{14}N}(\alpha ,p)\mathrm{^{17}O}$ reaction rate exceeds the rate of $^{17}$O burn up through conventional $\mathrm{^{17}O}(p,\alpha)\mathrm{^{14}N}$ and $\mathrm{^{17}O}(p,\gamma)\mathrm{^{18}F}$ processes. It is shown that these factors can enhance the $^{17}$O abundance in the core as compared with standard estimates. For the steady state case, the abundance enhancement is estimated to be as high as $\sim 10^2$ in the outer core region. A conjecture is made that other CNO suprathermal $(\alpha ,p)$ reactions may also alter abundances of CNO elements, including those generating solar neutrinos.
\end{abstract}


\maketitle

\section{\label{intro}Introduction}

A key issue for a proper description of nuclear burning processes in the Sun is an accurate treatment of chain reaction kinetics in the solar core plasma. The major nuclear inputs in kinetics simulations are rate parameters (reactivities) for reactions forming the $pp$ chain and carbon-nitrogen-oxygen (CNO) cycle in the Sun. It is known that standard solar models (SSMs) rely on a nuclear reaction network operating with thermal reactivities for particles having Maxwellian velocity distribution functions. At the same time, a question whether non-Maxwellian distortions of particle distributions may appear and affect reaction rates in the solar core has still been poorly studied. Clayton et al. \cite{clay74,clay75} analyzed how some non-Maxwellian distortions can alter the generation of solar neutrinos. In the context of the $^8$B neutrino problem, the authors introduced a radical \emph{ad hoc} assumption that ion distributions have depleted high-energy tails and thus depart from a Maxwellian function. This model \emph{decreases} the contribution to reaction rates of fast particles with energies large compared to the thermal energy. Although the model was shown to provide a pathway toward a solution of the neutrino problem, later Bahcall argued \cite{bahc89} that the degree of depletion needed is unlikely to occur in the dense core.

In the meantime, other non-Maxwellian phenomena which \emph{increase} the contribution of fast particles can manifest in the solar core plasma. These are suprathermal processes triggered by fast nucleons and lightest nuclei naturally appearing in the plasma. These particles are predominantly generated in exoergic reactions of the $pp$ chain and usually have MeV energies. When slowing down in the plasma they can undergo in-flight nuclear reactions and thus contribute to total reaction rates. The level of this suprathermal effect depends on a particular reaction. For exoergic processes, candidates most appropriate for the effect to manifest are reactions with pronounced resonances at suprathermal energies. In turn, direct reactions with cross sections moderately depending on energy are less sensitive to fast particles. As for endoergic processes, most of them can be appreciably influenced by fast particles having energies in excess of reaction threshold $E_\mathrm{thr}$.

One may expect therefore that energetic particles provide different impacts on forward (exoergic) $i+j \rightarrow k+l$ and reverse (endoergic) $k+l \rightarrow i+j$ reactions. This can lead to a deviation of the relation between forward and reverse reactivities, $\langle\sigma v \rangle_{ij \rightarrow kl}$ and $\langle\sigma v \rangle_{kl \rightarrow ij}$, from a standard law for Maxwellian plasma
\begin{equation}\label{eq:rev-for}
    \langle\sigma v \rangle_{kl \rightarrow ij} =
    \langle\sigma v \rangle_{ij \rightarrow kl} A_{ijkl} G_{ijkl} \exp(-Q/T).
\end{equation}
In Eq.~(\ref{eq:rev-for}), the quantity $Q$ is the Q-value of the forward reaction, $T$ is the plasma temperature in units of energy, $A_{ijkl}$ and $G_{ijkl}$ are algebraic combinations of particle mass numbers $A$ and temperature-dependent partition functions $G$, respectively (for details, see \cite{nacre99}). The violation of this relation caused by non-Maxwellian particles in astrophysical plasmas was recently obtained for several main processes of big bang nucleosynthesis \cite{voro11,voro12} and for some reactions of the solar CNO cycle \cite{voro15}. In particular, it was shown that 8.674-MeV $\alpha$ particles produced in the $\mathrm{^7Li}(p,\alpha)\alpha$ reaction of the $pp$ chain in the Sun crucially affect the balance of $p+\mathrm{^{17}O} \rightleftarrows \alpha + \mathrm{^{14}N}$ reactions, so that the reverse $(\alpha,p)$ process can even block the forward $(p,\alpha)$ one, distorting running of the CNO cycle.

This particular finding suggests the role of fast reaction-produced ions in the solar core may be unappreciated in SSMs, and in a greater or lesser degree it may concern processes of the $pp$ chain and CNO cycle. Furthermore, if we find that suprathermal reactions are capable of altering some element abundances in the core, suprathermal corrections to the fluxes of some solar neutrinos may also be obtained. In this context, it is worthwhile to note that the CNO cycle subdominant for energy generation in the Sun is an important source of neutrinos released in the $\beta^{+}$-decay of $^{13}$N, $^{15}$O, and also $^{17}$F nuclei. The CNO neutrinos carry valuable information on solar core metallicity \cite{haxt08}, and may also provide an independent test on the core temperature determined earlier on the basis of Sudbury Neutrino Observatory and Super-Kamiokande experiments.

All this serves as a clear argument in favor of a further study of non-Maxwellian (suprathermal) reactions in the solar interior. In this paper, a consistent model to properly describe these processes is formulated. The model is based on a formalism of in-flight reaction probability, operates with different approaches for the treatment of fast particle slowing-down in a matter, and takes into account some plasma peculiarities, such as the effect of electron degeneracy on in-flight reaction rates and electron screening of thermal reactions in the core. On the example of the CNO cycle we will demonstrate in detail how the model works and will improve results on $\alpha$-particle-induced suprathermal processes published previously \cite{voro15}.

\section{A model for non-Maxwellian processes in the solar core}
\label{model}

Let us consider non-relativistic fast particles $k$ produced in an exoergic reaction $i+j \rightarrow k + \cdots$ in the core plasma. To examine their effect on some $k+l$ reaction, one needs to know the particle emission rate $R_{k,ij}$, the particle energy loss rate $dE_k/dt$ in the matter, and the $k+l$ reaction cross section $\sigma_{kl}$ at thermal as well as suprathermal energies.

The particle emission rate $R_{k,ij}$ is determined by the following equations
\begin{equation}\label{eq:k-rate}
    R_{k,ij} = N_k \times R_{ij},
\end{equation}
\begin{equation}\label{eq:ij-rate}
    R_{ij} =
    (1+\delta_{ij})^{-1}
    n_i n_j \langle \sigma v \rangle_{ij},
\end{equation}
\begin{equation}\label{eq:ij-sigv}
    \langle \sigma v\rangle_{ij} =
    (n_i n_j)^{-1}
    \int\int f_i (\mathbf v_i) f_j(\mathbf v_j)
    \sigma (|\mathbf v_i-\mathbf v_j|)|\mathbf v_i-\mathbf v_j|
    \,d\mathbf v_i \,d\mathbf v_j.
\end{equation}
In these equations, $N_k$ is the number of particles $k$ produced per pair of $(ij)$, $R_{ij}$ is the $i+j$ reaction rate, $\langle \sigma v \rangle_{ij}$ and $\sigma$ are the $i+j$ reactivity and cross section, respectively, $n_a$ ($a = i,j$) is the number density of species $a$ with the velocity distribution function $f_a(\mathbf v_a)$ normalized to $n_a$.

We will use a formalism of in-flight reaction probability to describe the suprathermal $k+l$ reaction. This probability $W_{kl}$ satisfies the equation
\begin{equation}\label{eq:dW}
    dW_{kl} =
    (1 - W_{kl})\frac{n_l \sigma(E_k)}{(dE_k/dx)} dE_k,
\end{equation}
the solution of which is
\begin{equation}\label{eq:W}
    W_{kl}(E_{k,0} \rightarrow E_\text{th}) =
    1 - \exp
    \left [
       \int_{E_\text{th}}^{E_{k,0}}
       \left(
         \frac{2E_k}{m_k}
       \right)^{1/2}
       \frac{n_l \sigma (E_k)}{(dE_k/dt)}
       \, dE_k
    \right ].
\end{equation}
It presents the probability that the fast particle $k$ undergoes the in-flight $k+l$ reaction while slowing in the plasma from an initial energy $E_{k,0}$ down to the thermal energy $E_\text{th} = 3T/2$. In Eq.~(\ref{eq:W}), $E_k$ is the particle energy in the laboratory frame, $m_k$ is the particle mass, $\sigma$ is the reaction cross section, $n_l$ is the number density of target nuclei $l$, and $(dE_k/dt)$ is the particle energy loss rate.

A proper choice of this rate is a crucial point for accurate calculations of $W_{kl}$. In the solar core plasma fast particles lose energy through Coulomb elastic scattering (Coul) off background charged species, and through nuclear elastic scattering (NES) off ambient ions $i$ (neutrons can be ignored). Accordingly,
\begin{equation}\label{eq:loss}
    \left( \frac{dE_k}{dt} \right) =
    \left( \frac{dE_k}{dt} \right)_\text{Coul} +
    \left( \frac{dE_k}{dt} \right)_\text{NES}.
\end{equation}
The NES term in Eq.~(\ref{eq:loss}) can be presented in a form \cite{naka81}
\begin{eqnarray}\label{eq:nes}
    \left( \frac{dE_k}{dt} \right)_\text{NES} =
      & - &
    \sum_{i}
    \left(\frac{2E_k}{m_k}\right)^{1/2} n_i E_k
    \left(1-\frac{3T}{2E_k}\right) \nonumber \\
      & \times &
    \frac{4\pi m_k m_i}{(m_k + m_i)^2}
    \int_b^1 \sigma (E_k,\mu) (1-\mu)\,d\mu ,
\end{eqnarray}
where $\sigma (E_k,\mu)$ is the differential cross section for $k-i$ NES (allowing for the contribution of Coulomb-nuclear interference), $\mu$ is the cosine of scattering angle in the center-of-mass frame, and $b = -1$ (if $i \neq k$) or 0 (if $i = k$). In most cases, the NES contribution to $(dE_k/dt)$ is subdominant and the majority of particle energy loss comes from Coulomb scattering. Its rate can be written as
\begin{equation}\label{eq:ces}
    \left( \frac{dE_k}{dt} \right)_\text{Coul} =
    \left( \frac{dE_k}{dt} \right)_e +
    \sum_i \left( \frac{dE_k}{dt} \right)_i,
\end{equation}
where subscripts $e$ and $i$ stand for bulk electrons and ions, respectively.

To compose a detailed picture of particle thermalization in the sola core plasma, we will consider different models for Coulomb slowing-down process.

1. The first one is a standard binary-collision model with a Debye cut-off described by Sivukhin \cite{sivu66}. We will refer to it as SIV66. In this model, the partial terms $(dE_k/dt)_j$ ($j=e,i$) in Eq.~(\ref{eq:ces}) are given by
\begin{equation}\label{eq:ces1-a}
    \left( \frac{dE_k}{dt} \right)_j =
    - \frac{4\pi (Z_k Z_j)^2 e^4}{\left(2m_j T_j \right)^{1/2}}
    n_j \ln\Lambda_{kj} \frac{\Psi (x_j)}{x_j},
\end{equation}
\begin{equation}\label{eq:ces1-b}
    \Psi (x_j) = \text{erf}(x_j) - \frac{2}{\pi ^{1/2}}
    \left(1 + \frac{m_j}{m_k} \right)
    x_j \exp (-x_j^{2}),
\end{equation}
\begin{equation}\label{eq:ces1-c}
    x_j =
    \left(\frac{m_j}{m_k}\frac{E_k}{T_j}\right)^{1/2}.
\end{equation}
In this equations, $n_j$ and $T_j$ are the number density and the temperature of plasma species $j$, while $m_b$ and $Z_b$ ($b=j,k$) are the mass and the charge number of particle $b$. The quantity $\ln\Lambda_{kj}$ is the Coulomb logarithm treated in classical or quantum-mechanical approximations
\begin{equation}\label{eq:clog-a}
    \ln\Lambda_{kj} =
    \begin{cases}
      \ln \left(1 + \lambda^2_\text{D}/\rho_{\perp}^2\right)^{1/2},
      &\text{if } \rho_\perp > \lambdabar  \\
      \ln \left(2 \lambda_\text{D}/\lambdabar \right) - 1/2,
      &\text{if } \lambdabar > \rho_\perp
    \end{cases}
\end{equation}
where $\lambda_\text{D}$ is the Debye shielding length, $\rho_\perp$ is the impact parameter for $\pi /2$ deflection, and $\lambdabar$ is the de Broglie wavelength. They are
\begin{equation}\label{eq:clog-b}
    \frac{1}{\lambda^2_\text{D}} =
    \sum_{j = e,i} \frac{4\pi Z^2_j e^2 n_j}{T_j},
    \quad
    \rho_\perp =
    \frac{Z_k Z_j e^2}{m_r u^2},
    \quad
    \lambdabar =
    \frac{\hbar}{m_r u},
\end{equation}
where $m_r = m_k m_j/(m_k+m_j)$ and $u = |\mathbf v_k - \mathbf v_j| \simeq \sqrt{v^2_k+2T_j/m_j}$. Equations (\ref{eq:ces1-a})-(\ref{eq:ces1-c}) were derived under the assumption of Maxwellian velocity distribution functions for all plasma species $j$.

2. The second model developed by Kamelander \cite{kame86} is based on the Fokker-Planck collision theory. We will refer to it as KAM86. This model makes it possible to carry out an extended analysis of particle slowing-down in a dense matter. It operates with Maxwellian as well as Fermi-Dirac distribution functions, taking into account degeneracy of electron component. For the energy loss rate $(dE_k/dt)_j$ ($j=e,i$) one obtains \cite{kame86,naka09}
\begin{equation}\label{eq:ces2-a}
    \left( \frac{dE_k}{dt} \right)_j =
    - \frac{8\pi^2 (Z_k Z_j)^2 e^4 (2m_k)^{1/2}}{m_j E^{1/2}_k}
    \ln\Lambda_{kj} \times G_j(v_k).
\end{equation}
The quantity $G_j$ is a function of particle velocity $v_k$. It is given by
\begin{equation}\label{eq:ces2-b}
    G_j(v_k) =
    J_{j,2}
    \left[
      1-\frac{m_j}{3E_k} \frac{\left(J_{j,4}+J_{j,1} v^3_k\right)}{J_{j,2}}
    \right],
\end{equation}
where
\begin{equation}\label{eq:ces2-c1}
    J_{j,1}(v_k) =
    \int^\infty_{v_k} v_j f_j(v_j) \,dv_j,
\end{equation}
\begin{equation}\label{eq:ces2-c2}
    J_{j,2}(v_k) =
    \int^{v_k}_0 v^2_j f_j(v_j) \,dv_j,
\end{equation}
\begin{equation}\label{eq:ces2-c3}
    J_{j,4}(v_k) =
    \int^{v_k}_0 v^4_j f_j(v_j) \,dv_j,
\end{equation}
and $f_j(v_j)$ is the velocity distribution function of plasma species $j$. If all species are Maxwellian, the integrals $J_{j,m}$ ($m = 1,2,4$) are taken analytically and after some algebra Eq.~(\ref{eq:ces2-a}) can be reduced to the form of Eq.~(\ref{eq:ces1-a}) in which $\Psi (x_j)$ is replaced by $\widetilde{\Psi} (x_j)$
\begin{equation}\label{eq:ces2-d}
    \widetilde{\Psi} (x_j) =
    (1-T_j/E_k)
    \left[
      \text{erf}(x_j) - \frac{2}{\pi ^{1/2}}
      x_j \exp (-x_j^{2})
    \right].
\end{equation}

Under conditions typical of the solar core ($\rho \sim 150$~g/cm$^3$ and $T \sim 1.3$~keV) the electron temperature $T_e$ is close to the Fermi energy $E_\text{F} = (\hbar^2/2m_e)(3\pi^2 n_e)^{2/3}$. For example, in the inner core $T_e/E_\text{F} \sim 2.4$ and the plasma electrons are in a weakly degenerate state. To incorporate electron degeneracy in our model, we properly determine the electron distribution function $f_e(v_e)$ which obeys Fermi-Dirac statistics\footnote{Note that one more consequence of electron degeneracy is Pauli blocking -- a restriction of scattering collision between an electron and other particle \cite{sugi13}. However, it plays a subdominant role and is not considered here.}
\begin{equation}\label{eq:fe-a}
    f_e(v_e) =
    \frac{1}{4\pi^3}
    \left(\frac{m_e}{\hbar}\right)^3
    \left[
      \exp\left(\frac{m_e v^2_e}{2T_e} - \eta\right) + 1
    \right]^{-1}.
\end{equation}
In Eq.~(\ref{eq:fe-a}), the degeneracy parameter $\eta$ ($=\mu/T_e$ with $\mu$ being the chemical potential) is chosen to satisfy the normalization condition
\begin{equation}\label{eq:fe-b}
    \int^\infty_0
    f_e(v_e) 4\pi v^2_e \,dv_e =
    n_e.
\end{equation}
The Coulomb logarithm $\ln\Lambda_{kj}$ is also corrected to electron degeneracy by using an \emph{ad hoc} procedure \cite{brys74,brys75} of the replacement $T_e$ in Eq.~(\ref{eq:clog-a}) by $(T^2_e + T^2_\text{F})^{1/2}$ where $T_\text{F}$ is the Fermi temperature. Substituting now the distribution function $f_e(v_e)$, Eq.~(\ref{eq:fe-a}) into Eqs.~(\ref{eq:ces2-c1})-(\ref{eq:ces2-c3}), one can calculate the energy loss rate $(dE_k/dt)_e$, Eq.~(\ref{eq:ces2-a}), through $k-e$ collisions with degenerate electrons.

3. The third model for energy loss being considered was developed by Skupsky \cite{skup77}. We will refer to it as SKUP77. This model is based on a different energy loss mechanism. It was assumed that a charged particle $k$ induces an electric field in a matter, which acts back on this particle and ultimately decreases its kinetic energy. For this mechanism, the energy loss rate is expressed as
\begin{equation}\label{eq:ces3-a}
    \left( \frac{dE_k}{dt} \right) =
    - \frac{Z^2_k e^2}{2\pi^2}
    \int \frac{\mathbf{k}\cdot \mathbf{v}_k}{k^2}
    \frac{\epsilon_\text{Im}}{\epsilon^2_\text{Re} + \epsilon^2_\text{Im}}
    \,d\mathbf{k},
\end{equation}
where $\epsilon_\text{Re}$ and $\epsilon_\text{Im}$ are the real and imaginary part of the matter dielectric function $\epsilon(\mathbf{k},\mathbf{k}\cdot \mathbf{v}_k)$, respectively. Using the random-phase-approximation (RPA) form of the quantum-mechanical dielectric function, Skupsky \cite{skup77} obtained that the particle energy loss to plasma electrons of arbitrary degeneracy is given by
\begin{equation}\label{eq:ces3-b}
    \left( \frac{dE_k}{dt} \right)_e =
    - E_k n_e \frac{Z^2_k e^4}{T^{3/2}_e} \frac{(2m_e)^{1/2}}{m_k}
    \frac{4}{3}
    \left[
      \frac{\pi}{F_{1/2}(\eta)} \,\frac{1}{e^{-\eta} + 1}
    \right]
    \ln\Lambda_\text{RPA},
\end{equation}
where $F_{1/2}(\eta)$ is the Fermi integral to order 1/2
\begin{equation}\label{eq:fermi12}
    F_{1/2}(\eta) =
    \int^\infty_0
    \frac{x^{1/2} \,dx}{e^{x-\eta} + 1}.
\end{equation}
The quantity $\ln\Lambda_\text{RPA}$ in Eq.~(\ref{eq:ces3-b}) is a generalization of the classical Coulomb logarithm. It has the following form
\begin{equation}\label{eq:ces3-c}
    \ln\Lambda_\text{RPA} =
    (1 + e^{-\eta})
    \int^\infty_0
    \frac{k^3}{(k^2+k^2_0)^2}
    \left[
      \exp\left(\frac{\hbar^2 k^2}{8m_e T_e} - \eta\right) + 1
    \right]^{-1}
    dk,
\end{equation}
where $k^2_0 = k^2_\text{D} F'_{1/2}(\eta)/F_{1/2}(\eta)$ and $k^2_\text{D} = 4\pi n_e e^2/T_e$. We should note that this formula somewhat differs from $\ln\Lambda_\text{RPA}$ presented in \cite{skup77}. Equations (\ref{eq:ces3-b}) and (\ref{eq:ces3-c}) were derived for the case where the particle velocity $v_k$ is less than the average electron velocity $\langle v_e \rangle$. It was also assumed that $\epsilon_\text{Im} < \epsilon_\text{Re}$.

4. Edie \emph{et al}. \cite{edie13} recently proposed a reduced model for particle energy loss through scattering off plasma electrons. We will refer to it as EVRG13. The model conveniently interpolates between limiting cases for $(dE_k/dt)_e$ based on classical and quantum kinetic equations. It gives
\begin{equation}\label{eq:ces4-a}
    \left( \frac{dE_k}{dt} \right)_e =
    - \frac{4\pi Z^2_k e^4}{m_e v_k} n_e \ln\Lambda \times \Psi (x_0),
\end{equation}
where the function $\Psi (x_0)$ is described by Eq.~(\ref{eq:ces1-b}). The argument $x_0$ is
\begin{equation}\label{eq:ces4-b}
    x_0 =
    \left(\frac{m_e}{m_k}\frac{E_k}{T_e}\right)^{1/2}
    \left[
      \frac{\sqrt{\pi}}{2F_{1/2}(\eta) (1 + e^{-\eta})}
    \right]^{1/3}.
\end{equation}
The approximation formula for $\ln\Lambda$ has the following form
\begin{equation}\label{eq:ces4-c}
    \ln\Lambda =
    \frac{2m_e v^2_e}{\hbar \omega_e}
    \,\frac{0.321 + 0.259x^2_0 + 0.0707x^4_0 + 0.05x^6_0}
         {1 + 0.13x^2_0 + 0.05x^4_0},
\end{equation}
where $\omega_e = \sqrt{4\pi n_e e^2/m_e}$. For the electron velocity $v_e$ in Eq.~(\ref{eq:ces4-c}) we consider the most probable velocity $v'_e$ satisfying the equation $df_e(v_e)/dv_e = 0$. With the electron distribution function $f_e(v_e)$, Eq.~(\ref{eq:fe-a}), we found that in the solar core $v'_e \simeq \sqrt{2T_e/m_e}$.

SKUP77 and EVRG13 operate only with $k-e$ scattering. To obtain the total Coulomb energy loss for these models, we have also taken into account $k-i$ scattering for Maxwellian ions given by Eq.~(\ref{eq:ces1-a}).

Note that, since the above four models are based on different approaches, it becomes possible to give an extended description of particle slowing-down in the solar core plasma.

Now we introduce some other informative parameters characterizing fast particles in the solar core. The particle thermalization range $l_{k,\text{th}}$ and time $\tau_{k,\text{th}}$ in the plasma are
\begin{equation}\label{eq:th-range}
    l_{k,\text{th}}(E_{k,0} \rightarrow E_\text{th}) =
    \int_{E_\text{th}}^{E_{k,0}}
    -\frac{(2E_k/m_k)^{1/2} dE_k}{(dE_k/dt)},
\end{equation}
\begin{equation}\label{eq:th-time}
    \tau_{k,\text{th}}(E_{k,0} \rightarrow E_\text{th}) =
    \int_{E_\text{th}}^{E_{k,0}}
    -\frac{dE_k}{(dE_k/dt)}.
\end{equation}
Since $\tau_{k,\text{th}}$ reflects the particle ``lifetime'' in a suprathermal state, the number density of suprathermal particles approximately is $n_{k,\text{sprth}} = R_{k,ij} \times\tau_{k,\text{th}}$, where $R_{k,ij}$ is the particle emission rate in some $i+j$ reaction, Eq.~(\ref{eq:k-rate}). One more informative parameter is the effective temperature $T_{k,\text{eff}}$ of non-Maxwellian particles. It can be evaluated by equating the pressure of these particles having a slowing-down distribution function to the pressure of Maxwellian particles \cite{estr06}. This procedure gives
\begin{equation}\label{eq:Teff}
    T_{k,\text{eff}} =
    \frac{2I_4(v_c/v_{k,0})}{3I_2(v_c/v_{k,0})} E_{k,0},
      \quad
    I_n(a) \equiv
    \int_0^1
    \frac{x^n}{a^3 + x^3} \, dx,
\end{equation}
where $v_c$ is the crossover velocity defined in \cite{estr06}. This temperature determines the average kinetic energy of non-Maxwellian particles through a relation $\langle E_k \rangle = 3T_{k,\text{eff}}/2$.

The quantity of particular importance is the rate of suprathermal $k+l$ reactions induced by fast particles $k$ in the solar core. Using the in-flight reaction probability $W_{kl}$ defined by Eq.(\ref{eq:W}), this rate $R_{kl,\text{sprth}}$ can be evaluated as
\begin{equation}\label{eq:Rfast}
    R_{kl,\text{sprth}} =
    R_{k,ij} \times W_{kl}(E_{k,0} \rightarrow E_\text{th}),
\end{equation}
where $R_{k,ij}$ is the particle emission rate in the $i+j$ reaction. Equation (\ref{eq:Rfast}) can be presented in a standard form for reaction rate
\begin{equation}\label{eq:svfast-a}
    R_{kl,\text{sprth}} =
    n_{k,\text{sprth}} n_l
    \langle \sigma v \rangle_{kl,\text{sprth}},
\end{equation}
where $\langle\sigma v\rangle_{kl,\text{sprth}}$ denotes the effective suprathermal $k+l$ reactivity
\begin{equation}\label{eq:svfast-b}
    \langle\sigma v\rangle_{kl,\text{sprth}} =
    \frac{R_{k,ij} W_{kl}}{n_{k,\text{sprth}} n_l} =
    \frac{W_{kl}}{\tau_{k,\text{th}} n_l}.
\end{equation}
Although $\langle\sigma v\rangle_{kl,\text{sprth}}$ is not a conventional reactivity, it reasonably indicates the strength of the suprathermal $k+l$ reaction.

It is possible to generalize Eq.~(\ref{eq:Rfast}) to the case of suprathermal reactions induced by fast particles $k$ having a continuous source energy spectrum $S(E'_k)$. Such particles can be produced, e.g., in a reaction like $i+j \rightarrow k+k_1+k_2$. The generalized form of Eq.~(\ref{eq:Rfast}) is obtained by folding the in-flight reaction probability $W_{kl}$ over the source energy spectrum. If this spectrum  $S(E'_k)$ covers the range of energies $E_1 \leq E'_k \leq E_2$, the suprathermal reaction rate is
\begin{equation}\label{eq:Rfast-gen}
    R_{kl,\text{sprth}} =
    R_{k,ij} \times
    \left [
       \int_{E_1}^{E_2}
       S(E'_k) \,dE'_k
    \right ]^{-1}
    \int_{E_\text{m}}^{E_2}
     W_{kl}(E'_k \rightarrow E_\text{th}) S(E'_k) \,dE'_k,
\end{equation}
where $E_\text{m} = \max \,[E_1,E_\text{th}]$. Note that for monoenergetic particles with $S(E'_k) \sim \delta (E'_k-E_{k,0})$ this equation is reduced to Eq.~(\ref{eq:Rfast}).

We complete the description of our model by a method of accounting for electron screening of nuclear reactions in the solar core. As known, electron screening reduces the repulsive potential barrier between reacting ions and thereby enhances reaction cross sections $\sigma$ as compared with those $\sigma_\text{bare}$ for bare nuclei (for details, see a review paper \cite{adel11} and references therein). To incorporate this effect in the model, we use a weak-screening approximation \cite{salp54} properly describing electron screening for $a+b$ reactions with $Z_aZ_b \lesssim 10$ \cite{gruz98}. The respective enhancement factor $f_{ab} = \sigma /\sigma_\text{bare}$ is given by
\begin{equation}\label{eq:screen-a}
    f_{ab} =
    \exp
    \left(
      \frac{Z_a Z_b e^2}{T \lambda'_\text{D}}
    \right),
\end{equation}
where $T$ is the plasma temperature and $\lambda'_\text{D}$ is the Debye shielding length with a degeneracy correction
\begin{equation}\label{eq:debye-deg}
    \frac{1}{\lambda^{'2}_\text{D}} =
    \sum_{i} \frac{4\pi Z^2_i e^2 n_i}T +
    \frac{4\pi e^2 n_e}{T} \left(\frac{f'}{f}\right).
\end{equation}
The quantity $f'/f \simeq 0.92$ accounts for electron degeneracy in the solar core \cite{salp54}. Apparently, electron screening does not affect the probability of in-flight reaction $W_{kl}$, Eq.~(\ref{eq:W}), as this process is induced by energetic particles. Nevertheless, it can still enhance the in-flight reaction rate, Eq.~(\ref{eq:Rfast}), due to an increase of the particle emission rate $R_{k,ij}$ in a screened thermal $i+j$ reaction.

\section{\label{result}Numerical results and discussions}

In this section we explore capabilities of our model on the example of suprathermal effects in the solar CNO cycle triggered by MeV $\alpha$ particles produced in some $pp$ chain reactions.

We consider that plasma electrons and ions have equal temperature $T = T_e = T_i$. The radial profiles of temperature and element number densities in the solar core obtained \cite{ayukov} by running the MESA code \cite{mesa} are employed in the present work.

\begin{figure}
\begin{center}
\includegraphics{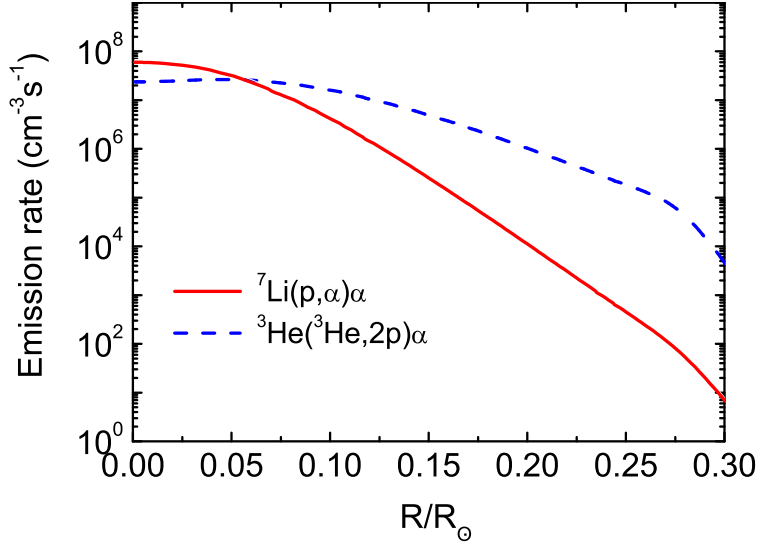}
\caption{\label{fig:a-rate}The emission rate $R_{\alpha,ij}$ of fast $\alpha$ particles in different regions of the solar core.}
\end{center}
\end{figure}

It was recently found \cite{voro15} that the major nuclear sources of MeV $\alpha$ particles in solar core are two reactions of the $pp$ chain
\begin{equation}\label{eq:p7li}
    p+\mathrm{^7Li} \rightarrow \alpha + \alpha
    \quad (Q = 17.348\text{ MeV}),
\end{equation}
\begin{equation}\label{eq:3he3he}
    \mathrm{^3He}+\mathrm{^3He} \rightarrow p + p + \alpha
    \quad (Q = 12.860\text{ MeV}).
\end{equation}
The $\alpha$-particle emission rates $R_{\alpha,ij}$ in these reactions are presented in Fig.~\ref{fig:a-rate}. Shown are the screened rates based on the corresponding reactivities for bare nuclei taken from the NACRE~II compilation \cite{nacre13}. The first reaction generates monochromatic $\alpha$ particles with energy $E_{\alpha,0}$ of 8.674~MeV, whereas the second one provides a continuous spectrum of $\alpha$ particles with energies up to 4.3~MeV. Let us calculate particle slowing-down characteristics introduced in Section \ref{model}, focusing mainly on more energetic $\alpha$ particles from the $\mathrm{^7Li}(p,\alpha)\alpha$ reaction.

\begin{figure}
\begin{center}
\includegraphics{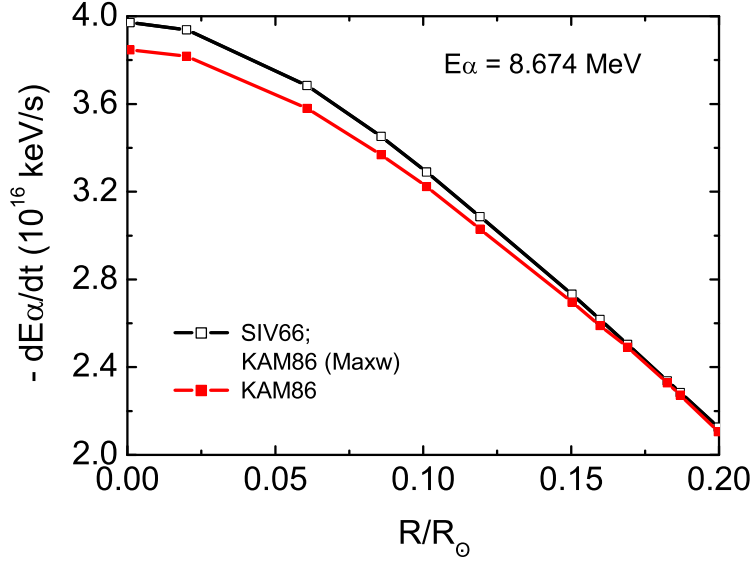}
\caption{\label{fig:eloss-a}The energy loss rate of a 8.674-MeV $\alpha$ particle from the $\mathrm{^7Li}(p,\alpha)\alpha$ reaction in the solar core plasma. The SIV66 and KAM86 (Maxw) models show the results for the Maxwellian plasma, while the KAM86 model takes into account the effect of electron degeneracy (see details in text).}
\end{center}
\end{figure}

\begin{figure}
\begin{center}
\includegraphics{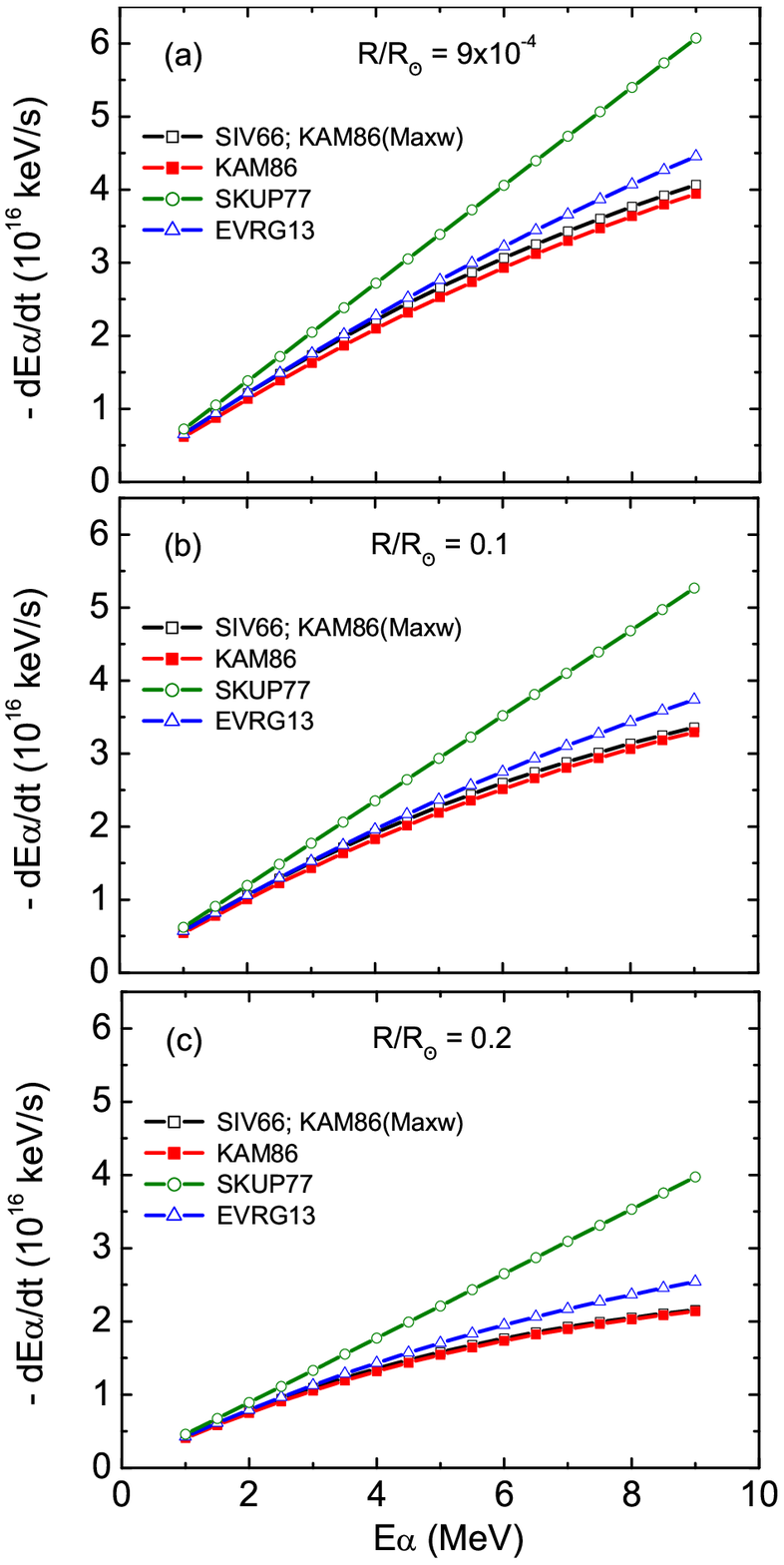}
\caption{\label{fig:eloss-b}The energy loss rate of an $\alpha$ particle as a function of its kinetic energy $E_\alpha$. Three regions of the solar core are considered: (a) inner core at $R = 9\times 10^{-4} R_\odot$; (b) middle core at $R = 0.1 R_\odot$; (c) outer core at $R = 0.2 R_\odot$. The curves marked with different symbols correspond to the different slowing-down models.}
\end{center}
\end{figure}

Figure \ref{fig:eloss-a} shows the energy loss rate of a 8.674-MeV $\alpha$ particle in the solar core plasma at $R \leq 0.2R_\odot$. To correctly clarify the role of electron degeneracy, we present here the results obtained within the \emph{single} slowing-down model, Eq.~\ref{eq:ces2-a}, considering first that all plasma species are Maxwellian and then taking into account degeneracy of the electron component. The corresponding curves in Fig.~\ref{fig:eloss-a} are denoted as KAM86 (Maxw) and KAM86. Due to spectral hardening of the Fermi-Dirac electron distribution, that is, an increase of the fraction of high-energy electrons as compared with the Maxwellian case, the $\alpha$ particle energy loss through $\alpha -e$ collision decreases. However, the reduction of $(dE_\alpha/dt)$ proves to be at a rather moderate level of not more than several percent as the plasma electrons are weakly degenerate. The energy loss rate found in the classical SIV66 model for Maxwellian plasma, Eq.~\ref{eq:ces1-a}, is also plotted in Fig.~\ref{fig:eloss-a}. This curve well coincides with that for Maxwellian KAM86.

\begin{figure}
\begin{center}
\includegraphics{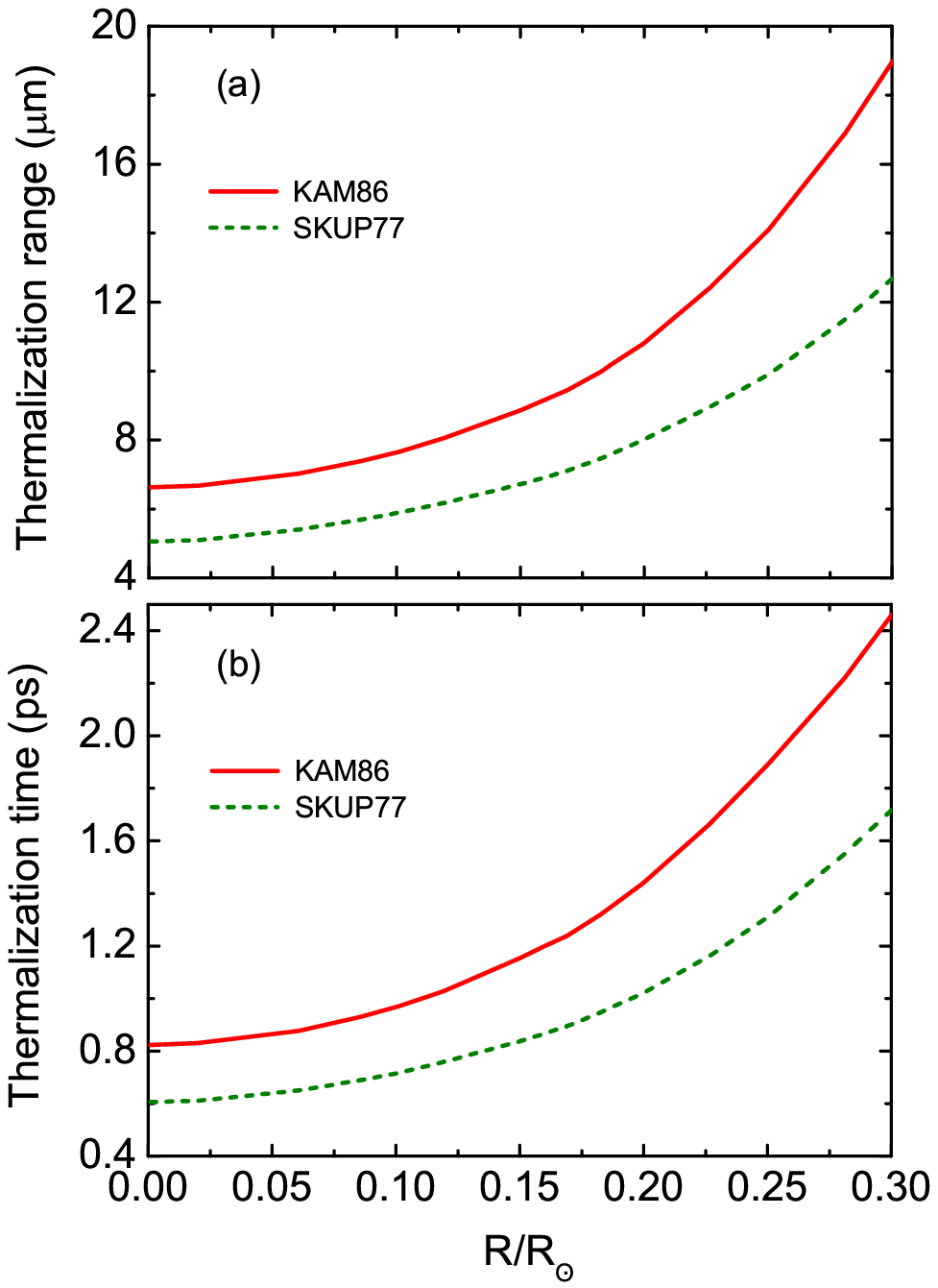}
\caption{\label{fig:therm}Parameters of 8.674-MeV $\alpha$ particle thermalization in the solar core. (a) Thermalization range $l_{\alpha,\text{th}}(E_{\alpha,0} \rightarrow E_\text{th})$. (b) Thermalization time $\tau_{\alpha,\text{th}}(E_{\alpha,0} \rightarrow E_\text{th})$.}
\end{center}
\end{figure}

Figure \ref{fig:eloss-b} shows the calculated energy loss rate $(dE_\alpha /dt)$ as a function of $\alpha$-particle energy $E_\alpha$ in the inner, middle, and outer core regions. All models being considered -- SIV66, KAM86 (Maxw) and KAM86, SKUP77, EVRG13 -- were used in these calculations. It is seen that both SIV66 and KAM86 models as well as the interpolated EVRG13 formula provide comparatively close values for $(dE_\alpha /dt)$ throughout the solar core. The lowest loss rate is predicted by the KAM86 model properly accounting for electron degeneracy. At the same time, the SKUP77 model, based on a completely different mechanism for energy loss, leads to a sizable increase of $(dE_\alpha /dt)$ at high energies. In view of this, we remind the reader that this model was developed for the case where the $\alpha$-particle velocity $v_\alpha$ is less than the average electron velocity $\langle v_e \rangle$. For 8.674-MeV $\alpha$ particles $v_\alpha /\langle v_e \rangle$ is close to unity, so one should bear in mind that some inaccuracy in the SKUP77 predictions at high energies is likely to occur. Nevertheless, the limiting cases of the lowest and highest particle energy loss correspond to the KAM86 and SKUP77 models, respectively.

\begin{figure}
\begin{center}
\includegraphics{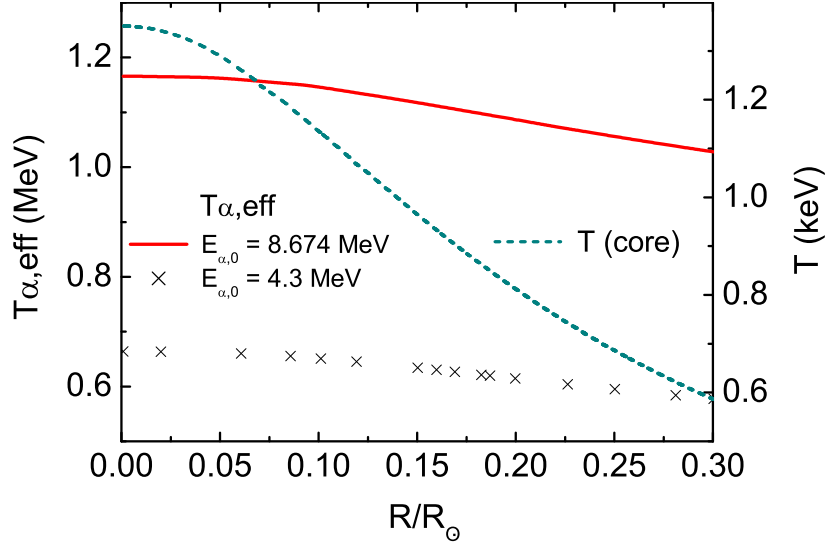}
\caption{\label{fig:temps}The effective temperature $T_{\alpha,\text{eff}}$ of reaction-produced $\alpha$ particles in a comparison with the solar core temperature $T$. The initial $\alpha$-particle energy $E_{\alpha,0} = 8.674$~MeV [the $\mathrm{^{7}Li}(p,\alpha)\alpha$ reaction] and 4.3~MeV [the upper energy for the $\mathrm{^{3}He}(\mathrm{^{3}He},2p)\alpha$ reaction].}
\end{center}
\end{figure}

The $\alpha$-particle thermalization range $l_{\alpha,\text{th}}$ and time $\tau_{\alpha,\text{th}}$, Eqs.~(\ref{eq:th-range}) and Eqs.~(\ref{eq:th-time}), for the limiting cases are presented in Fig.~\ref{fig:therm}. Both parameters increase towards the outer core in accordance with the decrease of $(dE_\alpha /dt)$ seen in Fig.~\ref{fig:eloss-b}. The $\alpha$-particle thermalization range $l_{\alpha,\text{th}} < 20$~$\mu$m, so all $R$-dependent plasma parameters $(T, n_e, n_i)$ entering Eq.~(\ref{eq:W}) can be assumed to be constant. This simplifies calculations of the in-flight reaction probability $W_{\alpha l}$. Figure \ref{fig:temps} shows the effective temperature  $T_{\alpha,\text{eff}}$, Eq.~(\ref{eq:Teff}), of reaction-produced $\alpha$ particles with the initial energy $E_{\alpha,0} = 8.674$~MeV [the $\mathrm{^{7}Li}(p,\alpha)\alpha$ reaction] and 4.3~MeV [the upper energy for the $\mathrm{^{3}He}(\mathrm{^{3}He},2p)\alpha$ reaction]. As seen, $T_{\alpha,\text{eff}}$ exceeds the core temperature $T$ by a factor up to 3 orders of magnitude.

\begin{figure}
\begin{center}
\includegraphics{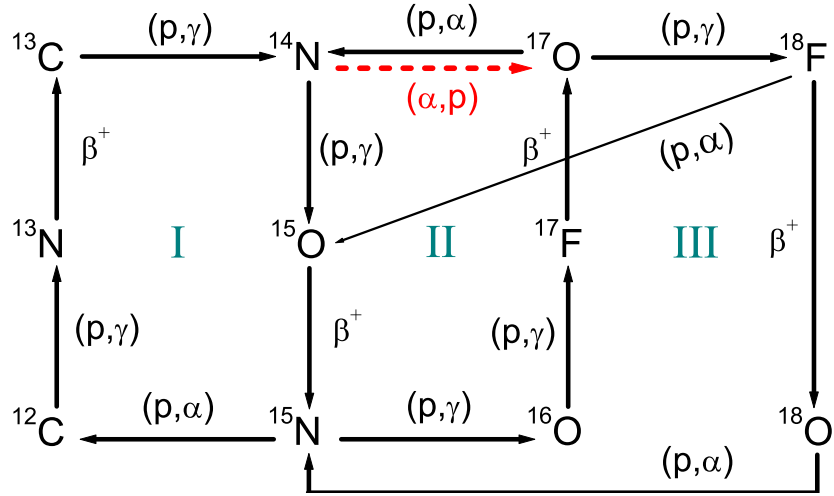}
\caption{\label{fig:CNO}The first three branches of the solar CNO cycle. The dashed arrow shows the endoergic reverse $\mathrm{^{14}N}(\alpha,p)\mathrm{^{17}O}$ reaction neglected in the SSM reaction network.}
\end{center}
\end{figure}

Thus, the main characteristics of fast $\alpha$ particles are evaluated and now one can examine a role which these particles play in the CNO cycle. The first three branches of the cycle are schematically shown in Fig.~\ref{fig:CNO}.  They involve several exoergic forward $\mathrm{A}(p,\alpha)\mathrm{B}$ reactions, whereas corresponding endoergic reverse $\mathrm{B}(\alpha,p)\mathrm{A}$ processes are neglected in the SSM reaction network. Thresholds $E_\text{thr}$ of the reverse $(\alpha ,p)$ processes are higher than 1~MeV, and according to Eq.~(\ref{eq:rev-for}) their Maxwellian reactivities $\langle\sigma v\rangle_{\alpha\mathrm{B}}$ in the solar core with the temperature of $\sim 1$~keV are dramatically suppressed by the factor $\exp(Q/T) \gtrsim \exp(1000)$. This means that Maxwellian $(\alpha ,p)$ nuclear flow does not appear in the CNO cycle. If however fast non-Maxwellian $\alpha$ particles are produced in the matter, the situation may change significantly.

We will show this on the example of the branch II (the CNO-II cycle), focusing on the following processes
\begin{equation}\label{eq:forward}
    p+\mathrm{^{17}O} \rightarrow \alpha + \mathrm{^{14}N} \quad (Q = 1.191\text{ MeV}),
\end{equation}
\begin{equation}\label{eq:reverse}
    \alpha + \mathrm{^{14}N} \rightarrow p+\mathrm{^{17}O} \quad (E_{\alpha,\text{thr}} = 1.531\text{ MeV}).
\end{equation}
The importance of the forward $(p,\alpha)$ reaction, Eq.~(\ref{eq:forward}), is that it closes the branch II and is one of main processes determining nuclear fusion rates in the CNO cycle \cite{adel98}. The reverse $(\alpha ,p)$ reaction, Eq.~\ref{eq:reverse}, neglected in SSM studies is shown by the dashed arrow in Fig.~\ref{fig:CNO}. For this reaction, the $\alpha$-particle threshold energy $E_{\alpha,\text{thr}}$ is only 1.531~MeV, so the process can easily be activated by 8.674-MeV $\alpha$ particles from the $\mathrm{^7Li}(p,\alpha)\alpha$ reaction. To calculate its characteristics, we used experimental data on the reaction cross section which exhibits a complicated resonant behavior (see, e.g., \cite{terw08} and references therein).

\begin{figure}
\begin{center}
\includegraphics{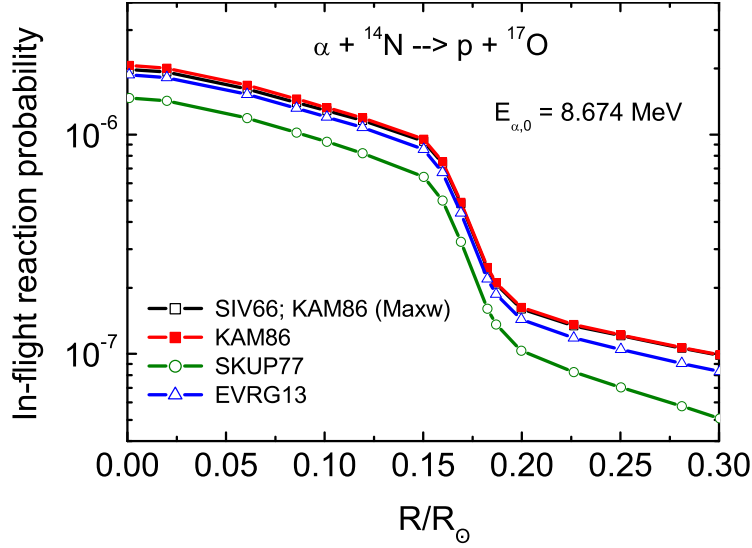}
\caption{\label{fig:rev-prob}The probability of the in-flight $\mathrm{^{14}N}(\alpha ,p)\mathrm{^{17}O}$ reaction in the solar core calculated for the different models of particle energy loss. The $\alpha$-particle initial energy $E_{\alpha ,0} = 8.674$~MeV.}
\end{center}
\end{figure}

\begin{figure}
\begin{center}
\includegraphics{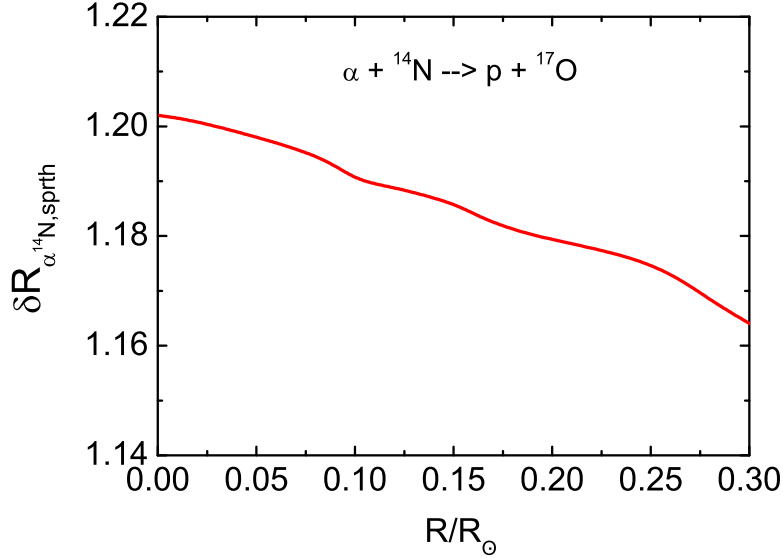}
\caption{\label{fig:ds-effect}The influence of electron degeneracy and electron screening on the suprathermal $\mathrm{^{14}N}(\alpha ,p)\mathrm{^{17}O}$ reaction rate $R_{\alpha \mathrm{^{14}N},\text{sprth}}$. Shown is the ratio $\delta R_{\alpha \mathrm{^{14}N},\text{sprth}} = R_{\alpha \mathrm{^{14}N},\text{sprth}} / R'_{\alpha \mathrm{^{14}N},\text{sprth}}$ (see details in text).}
\end{center}
\end{figure}

Figure~\ref{fig:rev-prob} shows the probability $W_{\alpha \mathrm{^{14}N}}$, Eq.~(\ref{eq:W}), for a 8.674-MeV $\alpha$ particle to undergo the in-flight $\mathrm{^{14}N}(\alpha ,p)\mathrm{^{17}O}$ reaction while slowing down in the solar core plasma. It is seen that all models for particle energy loss lead to a similar dependence of $W_{\alpha \mathrm{^{14}N}}$ on radius $R$. As one might expect, the highest and lowest probabilities are provided by the KAM86 and SKUP77 models, respectively.

Using Eq.~(\ref{eq:Rfast}), one can convert the in-flight reaction probability $W_{\alpha \mathrm{^{14}N}}$ to the corresponding suprathermal reaction rate
\begin{equation}\label{eq:a14Nrate}
    R_{\alpha \mathrm{^{14}N},\text{sprth}} =
    W_{\alpha \mathrm{^{14}N}} \times 2 \,n_p n_\mathrm{^{7}Li} \langle\sigma v\rangle_{p\mathrm{^{7}Li}\rightarrow 2\alpha} .
\end{equation}
Both electron degeneracy and electron screening in the plasma can enhance this rate. The first mechanism affects the in-flight probability $W_{\alpha \mathrm{^{14}N}}$ while the second one increases the thermal reactivity $\langle\sigma v\rangle_{p\mathrm{^{7}Li}\rightarrow 2\alpha}$ determining the fast $\alpha$-particle production. Their effect is clarified in Fig.~\ref{fig:ds-effect}. It shows the ratio $\delta R_{\alpha \mathrm{^{14}N},\text{sprth}}$ of the suprathermal rate $R_{\alpha \mathrm{^{14}N},\text{sprth}}$ (both mechanisms are included) to the rate $R'_{\alpha \mathrm{^{14}N},\text{sprth}}$ (both mechanisms are ignored). The rate enhancement by a factor of $\sim 1.2$ is observed at $R < 0.2 R_\odot$.

\begin{figure}
\begin{center}
\includegraphics{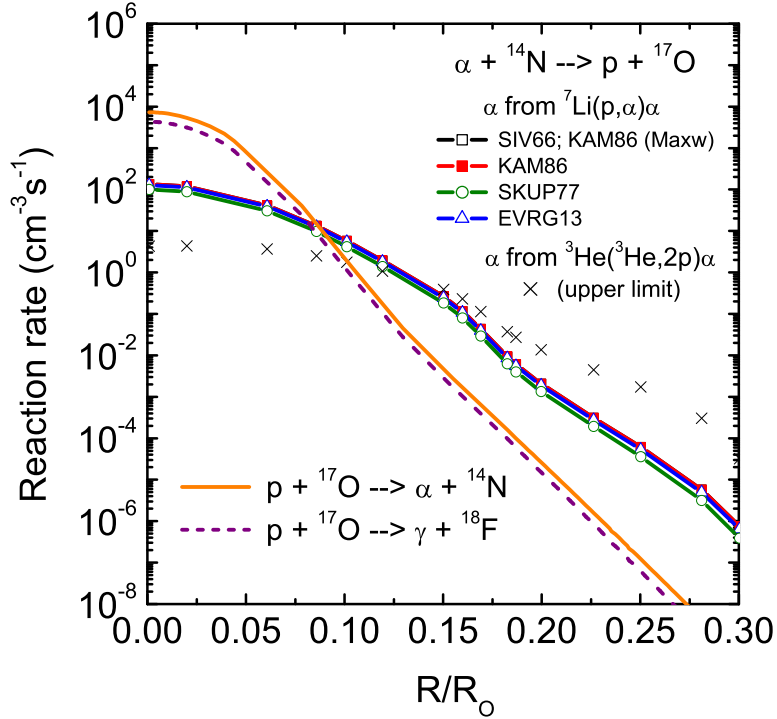}
\caption{\label{fig:rates}The suprathermal $\mathrm{^{14}N}(\alpha ,p)\mathrm{^{17}O}$ reaction triggered by fast $\alpha$ particles in a comparison with the $\mathrm{^{17}O}(p,\alpha)\mathrm{^{14}N}$ and $\mathrm{^{17}O}(p,\gamma)\mathrm{^{18}F}$ processes responsible for $^{17}$O burn-up in the solar core. Shown are the $\alpha +\mathrm{^{14}N}$ and $p+\mathrm{^{17}O}$ reaction rates determined by Eqs.~(\ref{eq:Rfast}) and (\ref{eq:ij-rate}), respectively. Two $\alpha$-particle sources are considered. The solid curves marked with symbols correspond to 8.674-MeV $\alpha$ particles from the $\mathrm{^{7}Li}(p,\alpha)\alpha$ reaction. Symbols ``$\times$'' present the illustrative results for $\alpha$ particles from the $\mathrm{^{3}He}(\mathrm{^{3}He},2p)\alpha$ reaction having the upper energy of 4.3~MeV.}
\end{center}
\end{figure}

Figure~\ref{fig:rates} shows the rate $R_{\alpha \mathrm{^{14}N},\text{sprth}}$ in a comparison with the rate of the forward $\mathrm{^{17}O}(p,\alpha)\mathrm{^{14}N}$ reaction. For completeness, the $\mathrm{^{17}O}(p,\gamma)\mathrm{^{18}F}$ reaction is also shown in this figure. Note that both $p + \mathrm{^{17}O}$ processes are responsible for $^{17}$O burn-up in the CNO cycle. Shown are the screened reaction rates. The $p + \mathrm{^{17}O}$ reactivities for bare nuclei were taken from a recent compilation \cite{ilia10} for temperatures above 10$^7$~K and extrapolated to lower temperatures $\sim 8\times 10^6$~K typical of the outer core region, taking into account the $T$-dependence of $\langle\sigma v\rangle$ obtained in \cite{nacre99}.

In Fig.~\ref{fig:rates}, the rate $R_{\alpha \mathrm{^{14}N},\text{sprth}}$ provided by 8.674-MeV $\alpha$ particles from the $\mathrm{^7Li}(p,\alpha)\alpha$ reaction is shown by solid curves marked with symbols (they are not resolved well on a logarithmic scale). Additionally, we present some estimates for another source of fast $\alpha$ particles (see Fig.~\ref{fig:a-rate}). It is the $\mathrm{^{3}He}(\mathrm{^{3}He},2p)\alpha$ reaction generating $\alpha$ particles in the 0--4.3~MeV range. The most effective energy for this reaction, i.e, the Gamow peak energy is $E_\text{G} \simeq 18.04 \,T^{2/3}$ \cite{clayton}, so in the solar core region $E_\text{G}$ varies within 14--22~keV. For such deep subbarrier energies reliable data on the $\mathrm{^{3}He}+\mathrm{^{3}He}$ $\alpha$-particle spectrum are not available in the literature that greatly complicates an analysis of the $\alpha$-particle contribution to the $\mathrm{^{14}N}(\alpha ,p)\mathrm{^{17}O}$ rate. However, some useful estimates can be done if we consider that all $\alpha$ particles from this reaction have the upper energy of 4.3~MeV. Although this case is not realistic, it demonstrates the upper limit for the $\mathrm{^{3}He}(\mathrm{^{3}He},2p)\alpha$ contribution. The respective results are shown by symbols ``$\times$'' (for the effective temperature $T_{\alpha ,\text{eff}}$ of 4.3-MeV $\alpha$ particles, see Fig.~\ref{fig:temps}). We consider these estimates as an argument in favor of a further study of the actual role of this reaction.

Figures \ref{fig:rev-prob} and \ref{fig:rates} demonstrate that the non-Maxwellian nuclear effects predicted on the basis of a simplified model \cite{voro15} can really occur in the solar core. One can schematically divide the core into three regions. In the inner core at $R < 0.1 R_\odot$ the reverse $\mathrm{^{14}N}(\alpha ,p)\mathrm{^{17}O}$ reaction is weaker than the forward $\mathrm{^{17}O}(p,\alpha)\mathrm{^{14}N}$ one and does not perturb the CNO running. In the narrow shell at $R = 0.087\text{--}0.091 R_\odot$ these two reactions have nearly equal rates, so the resultant nuclear flow between $^{17}$O and $^{14}$N almost vanishes, making the branch II unclosed (see Fig.~\ref{fig:CNO}). In the outer core at $R > 0.1 R_\odot$ the $\mathrm{^{14}N}(\alpha ,p)\mathrm{^{17}O}$ reaction enhanced by MeV $\alpha$ particles becomes much stronger than the $\mathrm{^{17}O}(p,\alpha)\mathrm{^{14}N}$ process. Accordingly, counter-clockwise nuclear flow in the branch II is redirected at $^{17}$O to clockwise flow in the branch III. It is worth noting that the region of this distortion covers $\sim 95$\% of the core with the radius $R \simeq 0.25 R_\odot$.

\begin{figure}
\begin{center}
\includegraphics{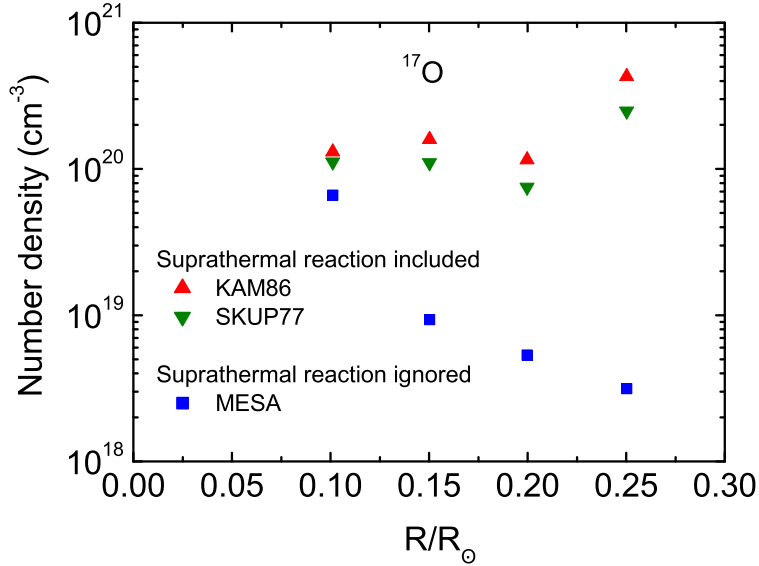}
\caption{\label{fig:17O}The influence of the suprathermal $\mathrm{^{14}N}(\alpha ,p)\mathrm{^{17}O}$ reaction on the $^{17}$O number density at $R = 0.1\text{--}0.25 R_\odot$.}
\end{center}
\end{figure}

One of possible consequences of this phenomenon is an increase of the $^{17}$O abundance in the outer core. Indeed, in this region the rate of $^{17}$O built-up through the $\mathrm{^{14}N}(\alpha ,p)\mathrm{^{17}O}$ reaction exceeds the rates of two competing $\mathrm{^{17}O}(p,\alpha)\mathrm{^{14}N}$ and $\mathrm{^{17}O}(p,\gamma)\mathrm{^{18}F}$ processes. It follows from Fig.~\ref{fig:CNO} that the $^{17}$O number density satisfies a rate equation
\begin{eqnarray}\label{eq:17O-rate}
    \frac{dn_\mathrm{^{17}O}}{dt} =
      & - &
    n_p n_\mathrm{^{17}O}
             \langle \sigma v \rangle_{p\mathrm{^{17}O}\rightarrow \alpha\mathrm{^{14}N}} -
    n_p n_\mathrm{^{17}O}
             \langle \sigma v \rangle_{p\mathrm{^{17}O}\rightarrow \gamma\mathrm{^{18}F}}
    \nonumber \\
      & + &
    \frac{n_\mathrm{^{17}F}}{\tau}
    + 2W_{\alpha \mathrm{^{14}N}} \,n_p n_\mathrm{^{7}Li} \langle\sigma v\rangle_{p\mathrm{^{7}Li}\rightarrow 2\alpha},
\end{eqnarray}
where $\tau = 93.04$~s is the mean lifetime of $^{17}$F. The last term in Eq.~(\ref{eq:17O-rate}) accounts for the suprathermal built-up of $^{17}$O. We estimate its contribution on the example of the rate equation nearly at steady state. In this case, the number density $n_\mathrm{^{17}O}$ is related to the thermal number density $n_{\mathrm{^{17}O},\text{th}}$ (obtained by ignoring the last term  in Eq.~(\ref{eq:17O-rate})) as
\begin{equation}\label{eq:sprth-th}
    n_\mathrm{^{17}O} \simeq
    n_{\mathrm{^{17}O},\text{th}} +
    \frac
      {2n_\mathrm{^{7}Li} W_{\alpha \mathrm{^{14}N}} \langle\sigma v\rangle_{p\mathrm{^{7}Li}\rightarrow 2\alpha}}
      {\langle \sigma v \rangle_{p\mathrm{^{17}O}\rightarrow \alpha\mathrm{^{14}N}} +
       \langle \sigma v \rangle_{p\mathrm{^{17}O}\rightarrow \gamma\mathrm{^{18}F}}}.
\end{equation}
These number densities are shown in Fig.~\ref{fig:17O}. It is seen that the suprathermal $\mathrm{^{14}N}(\alpha ,p)\mathrm{^{17}O}$ reaction is capable of essentially enhancing the $^{17}$O abundance in the outer core. The exact degree of this enhancement however may differ from that displayed in Fig.~\ref{fig:17O} because $^{17}$O may burn not in equilibrium. Nevertheless, these estimates suggest that the $^{17}$O abundance predicted by SSMs is likely to be underestimated in the outer core region.

\section{\label{conc}Conclusions}

In this paper, we have formulated the consistent model for the description of non-Maxwellian nuclear processes triggered by fast reaction-produced particles in the solar interior.  It is based on the formalism of in-flight reaction probability, operates with different methods for the treatment of particle slowing-down in the matter, and takes into account some peculiarities typical of nuclear processes in dense plasmas. These are the influence of electron degeneracy both on charged particle energy loss and suprathermal reaction rates, and electron screening of thermal processes in the solar core plasma. Our model extends the previous approach to study suprathermal solar reactions discussed in \cite{voro15} and refines its predictions.

To explore capabilities of this model, it has been applied to calculate the main characteristics of non-Maxwellian 8.674-MeV $\alpha$ particles generated in the $\mathrm{^{7}Li}(p,\alpha)\alpha$ reaction of the $pp$ chain, and to examine suprathermal processes in the solar CNO cycle induced by them. We particularly focused on the $\mathrm{^{14}N}(\alpha ,p)\mathrm{^{17}O}$ reaction neglected in SSM studies. The effect of electron degeneracy and electron screening on its rate in the solar core has been clarified. It has been shown that they increase the suprathermal $(\alpha ,p)$ rate by $\sim 20$\% at $R < 0.2 R_\odot$. The point of particular importance is that at $R > 0.1R_\odot$ this rate appreciably exceeds the rates of two conventional reactions -- $\mathrm{^{17}O}(p,\alpha)\mathrm{^{14}N}$ and $\mathrm{^{17}O}(p,\gamma)\mathrm{^{18}F}$. The first one closes the CNO-II cycle and both of them burn up $^{17}$O in the core.

This distorts running of the CNO cycle so that normal branching $\mathrm{^{14}N} \leftarrow \mathrm{^{17}O} \rightarrow \mathrm{^{18}F}$ of nuclear flow  (closing the branch II and starting the branch III) transforms to abnormal sequential flow $\mathrm{^{14}N} \rightarrow \mathrm{^{17}O} \rightarrow \mathrm{^{18}F}$. The region of this distortion covers $\sim 95$\% of the solar core volume. As a result, the $\mathrm{^{14}N}(\alpha ,p)\mathrm{^{17}O}$ reaction causes an enhancement of the $^{17}$O abundance in the core as compared with standard estimates. For the steady state case, the enhancement factor in the outer core reaches $\sim 10^2$. Rough estimates also suggest that an additional effect here may come from fast $\alpha$ particles produced in the $\mathrm{^3He}(\mathrm{^3He},2p)\alpha$ reaction of the $pp$ chain. Furthermore, we expect that suprathermal $(\alpha ,p)$ reactions other than $\mathrm{^{14}N}(\alpha ,p)\mathrm{^{17}O}$ may also alter abundances of CNO elements, including those generating solar neutrinos. This issue however is addressed to a future publication.

In closing, the applicability of the model presented is not restricted to solar studies and it can be used for an analysis of non-Maxwellian processes in some other stars, if main sources of energetic particles are accurately identified.

\end{document}